\documentclass[usegraphicx,usenatbib]{mn2e1}

\usepackage{epsfig}
\usepackage{longtable}
\usepackage{amssymb}
\usepackage{amsmath}

\begin{document}

\title[Velocity dispersion measurements of dwarf galaxies in the Coma cluster]{Velocity 
dispersion measurements of dwarf galaxies in the Coma cluster - implications for the 
structure of the fundamental plane}
\author[Cody et al.]
       {A.M. Cody$^{1,2}$, D. Carter$^3$, T.J. Bridges$^{1,4}$, 
       B. Mobasher$^{5,7}$, B.M. Poggianti$^6$. 
       \\$^1$ Anglo-Australian Observatory, PO Box 296, Epping NSW 1710, Australia.
       \\$^2$ Astronomy Department, Caltech, MC 105-24, 1201 East California Blvd, 
       Pasadena CA 91125, USA. 
       \\$^3$ Astrophysics Research Institute,
       Liverpool John Moores University, Twelve Quays House, Egerton
       Wharf, Birkenhead CH41 1LD, UK.
       \\$^4$ Department of Physics, Engineering Physics, and Astronomy, 
       Queen's University, Kingston, Ontario, K7L 3N6, Canada.
       \\$^5$ University of California, Riverside, 900 University Ave., Riverside, CA 
       92521, USA.
       \\$^6$ Osservatorio Astronomico di INAF, Universit\`{a} di Padova, 
       Vicolo dell'Osservatorio 5, I-35122 Padova, Italy.
       \\$^7$ Visiting Astronomer, Kitt Peak National Observatory, National 
       Optical Astronomy Observatories, which is operated by the \\
       \ Association of Universities for Research in Astronomy, Inc. (AURA) under 
       cooperative agreement with the National Science \\ \ Foundation}
\date{Accepted ... . Received ... ; in original form ... }
\pagerange{\pageref{firstpage}--\pageref{lastpage}}
\pubyear{2008} \volume{000} \pagerange{1} \onecolumn

\maketitle \label{firstpage}

\begin{abstract}

We present intermediate-resolution spectroscopic data for a set of dwarf and giant galaxies 
in the Coma Cluster, with $-20.6 < M_R < -15.7.$ The photometric and kinematic properties of 
the brighter galaxies can be cast in terms of parameters which present little scatter with 
respect to a set of scaling relations known as the Fundamental Plane. To determine the form 
of these fundamental scaling relations at lower luminosities, we have measured velocity 
dispersions for a sample comprising 69 galaxies on the border of the dwarf and giant regime. 
Combining these data with our photometric survey, we find a tight correlation of luminosity 
and velocity dispersion, $L \propto \sigma^{2.0}$, substantially flatter than the 
Faber-Jackson relation characterising giant elliptical galaxies. In addition, the variation 
of mass-to-light ratio with velocity dispersion is quite weak in our dwarf sample: ${M/L} 
\propto \sigma^{0.2}.$ Our overall results are consistent with theoretical models invoking 
large-scale mass removal and subsequent structural readjustment, e.g., as a result of 
galactic winds. 
\end{abstract}


\section{Introduction}
\label{introduction}

Dwarf galaxies ($M_R > -17.5$) are an important constituent of the Universe. They 
outnumber normal and giant galaxies, and form a distinct family of objects with very 
different fundamental properties from spirals and ellipticals (Kormendy 1985; Ferguson 
\& Binggeli 1994). The form of the low-mass end of the galaxy mass distribution is an 
important diagnostic of galaxy formation theories (White \& Frenk 1991). A clear 
understanding of the properties of dwarfs is essential to explore their relation to 
giant galaxies, to test galaxy formation models, and to establish a local calibrating 
sample to study properties of low-luminosity galaxies at high redshift. Since the dwarfs 
are likely to contain large amounts of dark matter in their haloes (Aaronson 1983; Mateo 
1998; Wilkinson {\sl et al.} 2002), any study of their evolution must also take account 
of their internal dynamics.

The fundamental plane (FP, Djorgovski \& Davis 1987; Dressler {\sl et al.} 1987) 
relating luminosity, velocity dispersion, surface brightness, and scale length for galaxies 
presents an important tool for the study of the dwarf galaxy population. In clusters more 
distant than Virgo and Fornax, spatially resolved spectroscopy becomes very difficult, but it 
is still possible to measure integrated or central velocity dispersions of large samples with 
multi-object fibre fed spectrographs. Thus, it is now becoming feasible to examine how the 
kinematics of low surface brightness galaxies fit in with those of the better-studied giants.

There are serious questions over whether dwarfs follow a continuous sequence with brighter 
ellipticals in their photometric and kinematic parameters.  Nieto {\sl et al.} (1990) were among 
the first to extend the FP to galaxies with low luminosity and mass. From a sample of 17 galaxies 
with $M_B$ in the range -19.4 to -15.3, they found that dwarf ellipticals form a low mass 
extension to the FP, with a scatter too large to be accounted for by measurement errors alone. 
They also found that the halo globular clusters lie on the faintward extension of the FP. Held 
{\sl et al.} 
(1992, 1997) reached similar conclusions with a set of dwarf galaxies that excluded the two 
``most extreme'' local group dwarf spheroidals, which are observed to have very high 
mass-to-light ratios. Incorporating galaxies with luminosities intermediate between the local 
group dwarfs and giant ellipticals, they found evidence for a continuous trend linking all the 
objects in their sample on the FP. Peterson \& Caldwell (1993), on the other hand, assert that 
the fundamental plane for dwarf ellipticals is entirely different from that of the giants. 
Compiling a sample of strongly nucleated galaxies in the range $-17.8 < M_V < -16.1$ from the 
literature, they found a steeper dependence of luminosity on velocity dispersion and a change in 
mass-to-light ratio with luminosity.  They claim that this supports scaling relations predicted 
by Dekel \& Silk (1986, hereafter DS86) for the removal of interstellar gas by supernova driven 
winds in dwarf galaxies surrounded by dominant dark haloes. Nevertheless, their conclusions are 
heavily influenced by the inclusion of all the local group dwarf spheroidals; it remains to be 
seen whether these well represent the fundamental plane at faint magnitudes, and if the 
nucleation has substantially affected the results (cf.\ Geha, Guhathakurta \& van der Marel 
2002). Other studies have hypothesized that a mixture of formation mechanisms, including ram 
pressure stripping of gas-rich dwarf irregulars (van Zee, Skillman \& Haynes 2004), primordial 
formation followed by gas loss in a supernova driven wind (DS86; Arimoto \& Yoshii 1987; Yoshii 
\& Arimoto 1987, hereafter YA87), ``harassment'' (Moore, Katz \& Lake 1996), and tidal formation 
(Kroupa 1998), are involved in setting the observed galaxy properties.

More recently, Geha {\sl et al.} (2002) and Geha, Guhathakurta \& van der Marel (2003) have 
studied the FP for their sample of 17 Virgo dwarfs ($-17.52 < M_V < -15.48$), using the 
parameterisation of Bender, Burstein \& Faber (1992). They showed that the dwarf ellipticals 
occupy a plane parallel to, but offset from, normal ellipticals. Graham \& Guzm\'{a}n 
(2003), however, have argued for a continuous progression of FP parameters based on detailed 
surface photometry from HST archival images for a sample of 18 dE galaxies in the Coma 
cluster. They modeled the surface brightness profiles with the more general S\'{e}rsic 
(1968) relation rather than the de Vaucouleurs (1948) law to uncover a strong correlation 
between the S\'{e}rsic index $n$, and absolute magnitude. In particular, they find that 
dwarfs have a S\'{e}rsic index in the range 1--2, and in all cases much less than the de 
Vaucouleurs law. They contend that application of the de Vaucouleurs law where it does not 
fit will result in derivation of incorrect surface brightness and size parameters, and that 
these differences will affect lower luminosity galaxies more. They conclude that the 
photometric scaling relations are continuous and linear, and hence normal and dwarf 
ellipticals form a single family. However, this does not explain why the dwarfs lie in a 
different region of the FP, as shown by Geha {\sl et al.} (2003), who fit S\'{e}rsic 
profiles, and find indices in the range 1--2.

To extend the selection of data on intermediate-luminosity dwarf ellipticals and contribute to 
ongoing analysis of these objects, we are engaged in a major study of the properties of galaxies 
in the Coma cluster. From our deep photometric survey (Komiyama {\sl et al.} 2002, hereafter 
Kom02) we have constructed a spectroscopic sample (Mobasher {\sl et al.} 2001, hereafter Mob01) 
with well defined selection functions. This has been used to investigate the dependence of their 
stellar components upon galaxy luminosity (Poggianti {\sl et al.} 2001a), morphology (Poggianti 
{\sl et al.} 2001b), and environment (Carter {\sl et al.} 2002). Spectroscopic observations have 
been used to identify the cluster members and to investigate the dynamics within the clusters of 
the dwarf and giant populations (Edwards {\sl et al.} 2002) and to study the properties of 
post-starburst galaxies and the correlation between their position and cluster substructure 
(Poggianti {\sl et al.} 2004). Here we report on higher resolution spectroscopic observations of 
a subsample of these galaxies, in order to establish an unbiased FP. Studies of the faint 
extension of the FP, including ours, concentrate on galaxies in clusters. Largely this is for 
practical reasons: observations require long exposure times and to obtain sufficient samples 
multi-object spectrographs are required, which in turn require a high density of targets. While 
cluster galaxies are more affected by interactions with their environment than field galaxies, 
by studying regions of different density within the same cluster we can hope to quantify the 
effect of such interactions and to eliminate them as a source of uncertainty.

Two other recent studies address the same problem. The NOAO Fundamental Plane Survey
(NFPS, Smith {\sl et al.} 2004) is a large survey of over 4000 galaxies in 93 clusters.
This survey concentrates on normal elliptical and lenticular galaxies, rather than
the dwarfs. The survey was designed for clusters at a range of redshifts, and was carried
out at a resolution approximately a factor three worse than ours; thus it is most
reliable for velocity dispersions above 100~km~s$^{-1}$. Matkovi\'{c} \& Guzm\'{a}n
(2005, hereafter MG05) and  Matkovi\'{c} \& Guzm\'{a}n (2007) present velocity dispersion 
measurements for a sample of faint
early-type galaxies in the core of the Coma cluster. Again, their spectral resolution is
worse than that of this study, by a factor of 1.6. Although they present velocity
dispersions as low as 35~km~s$^{-1}$, they require a correction for systematic errors 
at this dispersion, and our sample is in general of galaxies of lower velocity 
dispersion. Furthermore, their sample is largely in the core of the cluster, whereas 
our sample contains both the core and lower density regions. Accordingly, our data put us in 
a position to provide a valuable extension to lower luminosity and less dense environment, with 
which to study the problem of the origin and properties of dwarf galaxies.


\section{Observations} 
\label{observations}

\subsection{Source Selection} 
\label{selection}

The spectroscopic targets for this study were selected from our wide-angle photometric (Kom02) 
and spectroscopic (Mob01), and subsequent papers in the series) catalogues of Coma cluster 
galaxies. The photometric catalogue provides B and R mags (to $R\sim 22$) and colours with an 
accuracy of 0.1 mag for three fields, covering 1.33 deg$^2$ in Coma. The spectroscopic 
catalogue covers the Coma1 (central) and Coma3 ($\sim$1 degree SW of the cluster centre, 
containing NGC 4839) fields, and was obtained using the Wide-field Fibre Optic Spectrograph 
(WYFFOS) on the William Herschel Telescope. The WYFFOS spectra, with a resolution of 6--9\AA, 
yield redshifts (and hence cluster membership) and several spectral line indices sensitive to 
stellar population ages and metallicities (e.g. Poggianti {\sl et al.} 2001a). The sample 
chosen for this present, higher-resolution spectroscopic study, was selected to include only 
spectroscopically confirmed members of the Coma cluster and to have $14.5 < R_{Kron} < 19.4$ 
and $19.5 < \langle\mu_R\rangle < 24.3$ mag/arcsec$^2$, where $R_{Kron}$ and 
$\langle\mu\rangle$ are the Kron magnitudes and average surface brightness over the Kron radius 
(an intensity weighted radius; Mob01), respectively. These limits correspond to $-20.6 < M_R < 
-15.7$, and thus our sample includes galaxies on either side of the boundary between dwarfs and 
giants, which we define following Mob01 to be at $M_R = -17.5$ or $R = 17.6$ at the distance of 
Coma, assuming a distance modulus of 35.1 for the cluster (Baum {\sl et al.} 1997). For 
simplicity, we will primarily refer to these objects as dwarfs, given that the sample as a 
whole covers a distinct region of parameter space in luminosity and velocity dispersion 
compared to previous studies. Similar numbers of galaxies were observed in the high density 
core of the cluster, and in the outskirts (the Coma3 region of Kom02), with a total of 70 
fibres on galaxies in the former and 65 in the later. The positional accuracy of the selected 
targets is $0\farcs5$, sufficient for spectroscopic purposes.

\subsection{Spectroscopic Observations}
\label{spectra}

The intermediate-resolution spectroscopic observations for the present study were obtained with 
the 3.5-m Wisconsin-Indiana-Yale-NOAO (WIYN) telescope, using the Hydra multifibre 
spectrograph. The targets were galaxies in the Coma1 and Coma3 fields, selected as described 
above. These fields were chosen to allow a large density contrast in the location of sample 
galaxies.  The observations were performed over three nights between 28 April and 1 May 2003. 
We used an exposure time of $24\times 30$ minutes and $14\times 30$ minutes for Coma~1 and 
Coma~3 fields, respectively. The spectra covered a range $\sim 4700-5700 \AA$ at $\Delta\lambda 
\sim 0.49 \AA$ per pixel and 1.2~$\AA$ spectral resolution, encompassing the region covered by 
Mg{\it b} and H$\beta$ lines. Copper-argon lamp exposures were taken for wavelength 
calibration, and dome and twilight flatfields and bias frames were obtained. During each 
observation, approximately 80 out of the 100 3$\arcsec$-diameter fibres were placed on target 
galaxies. The remaining fibres were assigned to sky. A number of standard stars with spectral 
types G8V--M0~III were also observed to provide velocity templates and calibration sources.

\section{Data Reduction}
\label{reduction}
\subsection{General procedures}
\label{general}
\begin{figure}
\begin{center}
{\includegraphics[width=12cm]{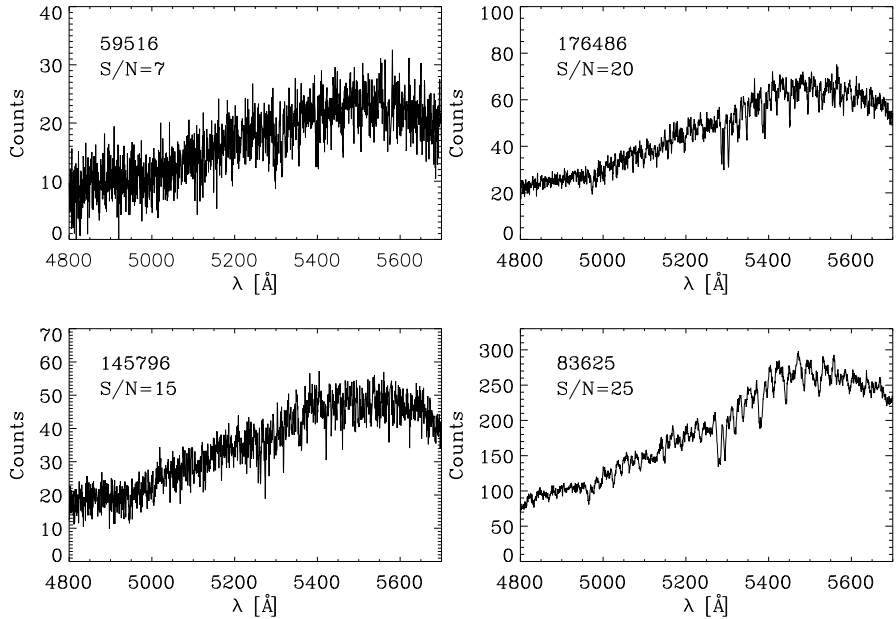}}
\end{center}
\caption{\label{spectrumplot}
Examples of spectra with various signal-to-noise values. The Komiyama identification 
number is shown (see Table~A1).
}
\end{figure}

Standard data reduction techniques were employed. All raw spectra were bias subtracted,
using a collection of zero frames that were median-combined with a cosmic-ray rejection
algorithm.  Sets of dome and twilight flatfield images taken at the beginning and end of
each night were combined in a similar manner. Arc lamp exposures taken in succession were
also combined.  Subsequent data reduction was performed with the IRAF DOHYDRA package.
Spectra were flatfielded, sky subtracted, wavelength calibrated, scattered light
subtracted, cleaned of bad pixels, and corrected for varying fibre throughput (using
flatfields). One night sky line, at 5577~$\AA$, partially remained and was masked out of later
spectral analysis. After object spectra were extracted, all frames taken of each field
were combined using median scaling and cosmic ray rejection. Sky subtraction was neglected
for the short exposures of bright standard stars. The final spectra were rebinned to a
log-linear wavelength scale with 2048 pixels. Signal-to-noise values were substantially
lower than expected (i.e., 60\% of the targets had signal-to-noise ratios of 10 per $\AA$ 
or lower), but none the less our spectral resolution still allows us to derive
velocities with reasonable uncertainties, as discussed in $\S$\ref{errors}. Several
example spectra are shown in Figure~\ref{spectrumplot}.

\subsection{Radial velocities and dispersions} 
\label{velocities}
To measure the redshift and amount of broadening present in each galaxy spectrum, we
adopted the Fourier cross-correlation technique, originally developed by Tonry \& Davis
(1979). We chose this over more sophisticated methods that provide second-order velocity
moments, because of the limitations of our mediocre signal-to-noise levels. Galaxies were
cross-correlated with each of five stellar velocity templates, using the IRAF package
FXCOR. All spectra were apodised down to 5\% at the ends with a cosine bell. In addition,
they were filtered in the fourier domain with a ramp function to remove additional unwanted
noise.

Galaxy spectra were cross-correlated with the template spectra in the region 4740 to 5726~$\AA$, with the 
night sky line masked out.  Four template stars-- HD~62509, HD~65583, HD~75839, and HD~90861-- were observed 
with the same spectroscopic setup and comprised spectral types G8V, K2~III, and K0~IIIb. HD~65583 was 
observed with two different fibres. Redshifts were computed from the position of cross-correlations peaks, 
and velocity dispersions were determined from the widths of the peaks.  In order to convert from full-width 
at half maximum (FWHM) of the cross-correlation peaks to a true velocity dispersion, calibration curves were 
produced by broadening the template stars with gaussian kernels of different velocities, and 
cross-correlating the resulting spectra against the original templates. This technique allows for reasonably 
accurate measurement of dispersions down to the instrumental resolution limit where the FWHM reaches 
$\sim$120~km~s$^{-1}$. At the distance of the Coma Cluster ($\sim$105~Mpc), the 3$\arcsec$ diameter aperture 
is equivalent to 1.53 kpc. Hence our velocity dispersions are effectively averaged out to a radius of 0.76 
kpc. We believe this is preferable to obtaining central velocity dispersions, since Geha {\sl et al.} (2002) 
have shown that galactic nuclei occupy a region of the fundamental plane separate from the underlying 
galaxy, and much closer to the globular clusters (a low-mass, high surface brightness region). To assess the 
degree of mismatch between template spectra and galaxies, we examined the spread of dispersion values 
obtained for each galaxy. No individual template produced consistently discrepant dispersions, and the 
1-$\sigma$ variation of values returned by different templates was typically 1-5~km~s$^{-1}$. Although the 
spread in the spectral types is small enough that perhaps little velocity variation would be expected, tests 
using more extensive template sets (see \S 4.1) did not produce systematic shifts in the results. Hence, for 
each galaxy we averaged the results of all templates. We used the Tonry-Davis $R$ coefficient (TDR; Tonry \& 
Davis, 1979) as a measure of goodness of fit to the cross-correlation peak; typically TDR values less than 
3.0 are unreliable. However, our adopted signal-to-noise requirement of S/N$>$7 (\S 3.3) proved to the 
dominant selection criterion, leading to the removal of all but one galaxy (GMP~4351) with TDR$<$6 from the 
sample. This remaining galaxy was retained since no single cross-correlation template produced a TDR value 
less than 3.0.

\subsection{Error analysis}
\label{errors}
\begin{figure}
\begin{center}
\rotatebox{270}{\includegraphics[width=9cm]{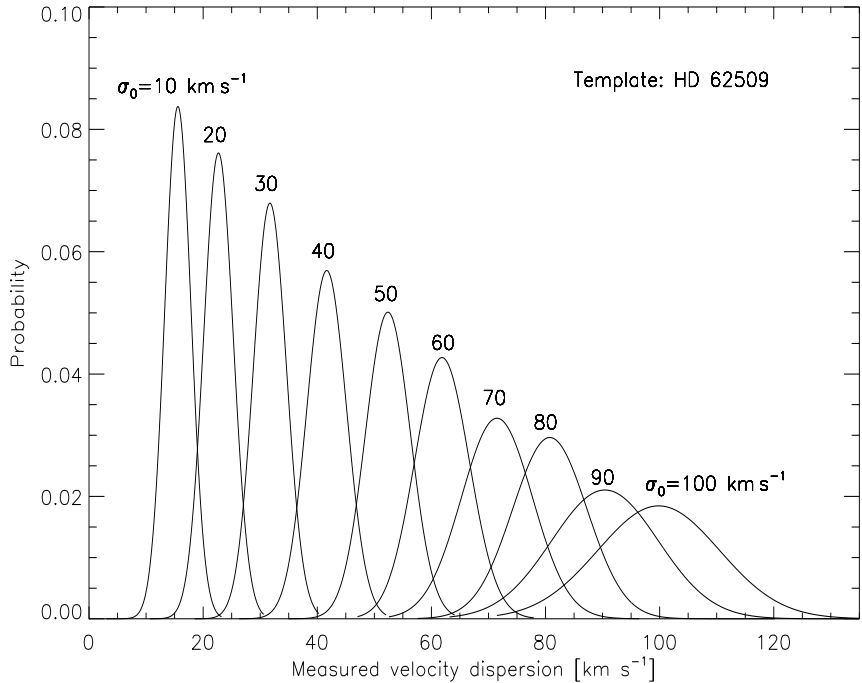}}
\end{center}
\caption{\label{fakegals}
Results of cross-correlating the spectrum of template star HD~62509 with fake galaxies
of S/N=15 per $\AA$ and velocity dispersions from $\sigma_0$=10 to 100~km~s$^{-1}$.
Each distribution of measured velocity dispersions shown is derived from a run of 300
simulations in which the template spectrum was cross-correlated against a fake galaxy
with a particular broadening $\sigma_0$ (as noted above each plotted distribution) and a
gaussian was fit to the results.
}
\end{figure}

The primary source of uncertainty in our results is poor signal-to-noise. We
suspect that misalignments of the fibres with the target galaxies due to a combination
of astrometric measurement and transformation and physical positioning
errors is responsible for this problem.  To determine the relationship between
signal-to-noise and accuracy of dispersion measurements, we have performed
bootstrap simulations in which template star spectra are broadened to a variety of
velocities, combined in equal proportions of spectral types to create a ``fake''
galaxy, and subsequently augmented with random noise to achieve a particular
signal-to-noise ratio (S/N). The resulting ``galaxy'' spectrum is then cross-correlated 
with the original templates to determine what the range of measured velocity dispersions 
would be. The process of adding noise and measuring the dispersion is carried out 300 
times for each S/N and velocity broadening value, and a gaussian is fit to the results.
An example simulation for the template star HD~62509 cross-correlated
against fake galaxies with S/N=15 per $\AA$ and broadening values, $\sigma_0$, from 10 to 
100~km~s$^{-1}$ is shown in Figure~\ref{fakegals}. Although each value of $\sigma_0$ 
leads to a symmetric, gaussian distribution of measured dispersion, in some cases there are
offsets between the value of $\sigma_0$ and the mean measurement. This occurs particularly
at low dispersions where the cross correlation method tends to produce overestimates. As
explained below, we include this effect as an additional source of uncertainty.
At higher dispersion, the larger broadening values also lead to a larger spread. Nevertheless, 
the widths and peaks of the distributions follow regular trends across the entire range
of $\sigma_0$, to which we have fit polynomial functions; this enables us to predict the 
distribution resulting from any $\sigma_0$.

\begin{figure}
\begin{center}
\rotatebox{270}{\includegraphics[width=9cm]{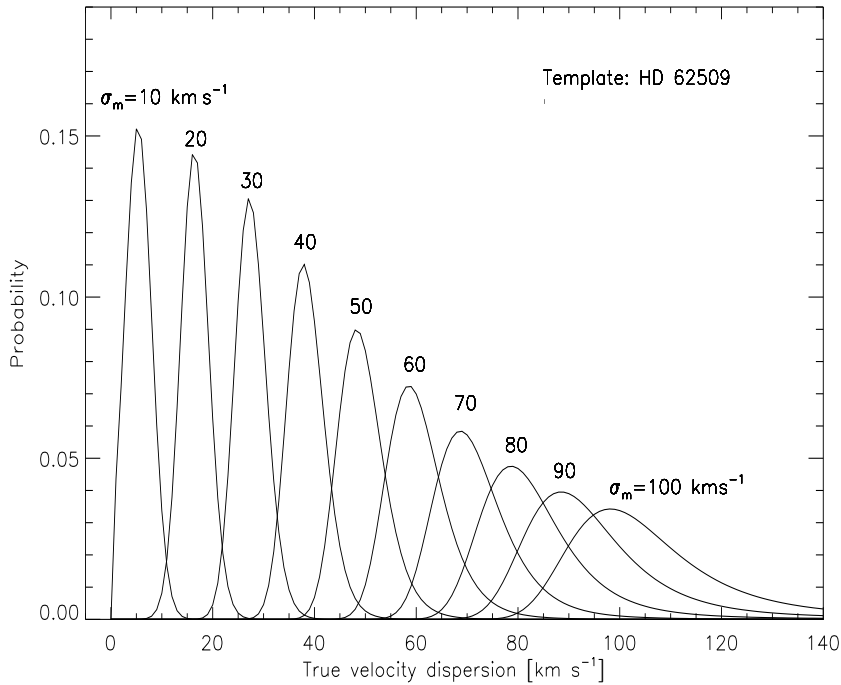}}
\end{center}
\caption{\label{fakegals1}The distributions in Figure \ref{fakegals},  
inverted to give the probability of true velocity dispersion as a function of 
the measured value (where the measured values, $\sigma_{\rm m}$, are plotted in increments 
of 10 here). 
}
\end{figure}

Calculating the uncertainties for a {\em measured} dispersion $\sigma_m$ requires an 
inversion of these distributions to obtain the probability that it was derived from any 
{\em true} dispersion, $\sigma_0$ (what is plotted in Figure~\ref{fakegals} shows the 
opposite). We have done this by discretising the probability distribution in 
increments of 1~km~s$^{-1}$ up to a maximum of 140~km~s$^{-1}$: for each 
$\sigma_{0,i}\in\{1,2,3,....139,140~{\rm km}~{\rm 
s}^{-1}\}$, the probability of measuring $\sigma_m$ is denoted 
P($\sigma_m|\sigma_{0,i}$). We solve for the values of P($\sigma_m|\sigma_{0,i}$) 
by integrating the gaussian distributions for each $\sigma_{0,i}$ from 
$\sigma_m$-0.5~km~s$^{-1}$ to $\sigma_m$+0.5~km~s$^{-1}$, since our measurements are 
probably not accurate to better than 1~km~s$^{-1}$. The conditional probability for 
true dispersion is hence given by: 
$$P(\sigma_{0,k}|\sigma_m)=P(\sigma_m|\sigma_{0,k})/\displaystyle\sum_{i=1}^{140} 
P(\sigma_m|\sigma_{0,i}).$$ The results of this computation are displayed in 
Figure~\ref{fakegals1}.

The resulting distributions of velocity dispersion illustrate how accurate our 
measurements are for low S/N. As seen in Figure~\ref{fakegals1}, they are 
slightly asymmetric, due to the fact that larger velocity dispersions produce a larger 
spread in the measured $\sigma$. Velocity dispersion is also correlated with galaxy 
brightness, in the sense that giant ellipticals have both larger velocity dispersions and 
luminosities than dwarfs. Therefore, because we observed the galaxies in the sample for 
equal periods of time, low dispersion values correlate with low S/N.

For each value of S/N, we combine the results of simulations involving each of the five
template stars. The 1-$\sigma$ uncertainties for each velocity dispersion value are derived
by integrating the distributions in Figure~\ref{fakegals1} to $\pm 34.1\%$ of their total
area to either side of the peak. If the peak does not fall on the measured value, the
difference between these two is incorporated (added in quadrature) as a systematic error.  
After adding the 1-$\sigma$ uncertainties from each template in quadrature, we find that we
can measure dispersions near 15~km~s$^{-1}$ with 25\% accuracy at S/N$\lesssim$7 per $\AA$
and near 25~km~s$^{-1}$ with 15\% accuracy for S/N$\sim$8. We attain 6\% accuracy for the
galaxies that have $\sigma$ $\sim$ 35~km~s$^{-1}$ and S/N$\sim$15. For S/N values above 20, and
dispersions larger than 40~km~s$^{-1}$, we achieve an uncertainty of 3\%. Because accurate
dispersion measurements require reasonably high signal-to-noise ratios, we have disregarded
most spectra with S/N$<7$. This condition was relaxed for several of the dwarfs with very low
velocity dispersion; errors of $\sim 25$\% were permitted for these objects, since we
believe it is important to extend the sample to galaxies with low surface brightness.

Uncertainties were calculated individually for each of the galaxies and are 
quite small-- on the order of 2-4~km~s$^{-1}$. 
The derived uncertainties could be overly optimistic-- especially given that our fake galaxy 
spectra consist of only five stars (of which one is the exact template), as opposed to 
millions in a true galaxy spectrum. We discuss this further in \S~4.1.
However, our confidence is increased by the fact that in most cases, the spread in velocity 
dispersion measurements from the five different template spectra is {\em less} than the 
uncertainty we have estimated from the simulations. The final tally of galaxies observed 
that we believe to have reliable velocity dispersion measurements includes 36 galaxies in 
Coma~1 and 33 galaxies in Coma~3.

\section{Results}
\label{results}
We present our measured velocities and velocity dispersions in Table~\ref{resultstable}. In 
this table column 1 gives the identification from Godwin, Metcalfe \& Peach (1983); column 2 
the identification from our own programme (Kom02, Mob01); column 3 a 
morphological type, estimated visually from the R band mosaic CCD image; columns 4 and 5 the 
J2000 position of the galaxy; columns 6 and 7 the R-band and B-R Kron magnitude (see Kom02
for a definition); column 8 the effective R-band surface brightness 
(average surface brightness within the effective radius; Kom02), columns 9 
and 10 the heliocentric recession velocity and its error in km~s$^{-1}$; columns 11 and 12 the 
velocity dispersion and its error in km~s$^{-1}$; and column 13 the signal-to-noise per $\AA$. 
In column 3, E and dE galaxies are delineated by the boundary $M_R = -17.5$ ($R = 17.6$ at the 
distance of Coma) chosen by Mob01.

\subsection{Comparison with external data}
\label{comparisons}

\begin{figure}
\begin{center}
{\includegraphics[width=12cm]{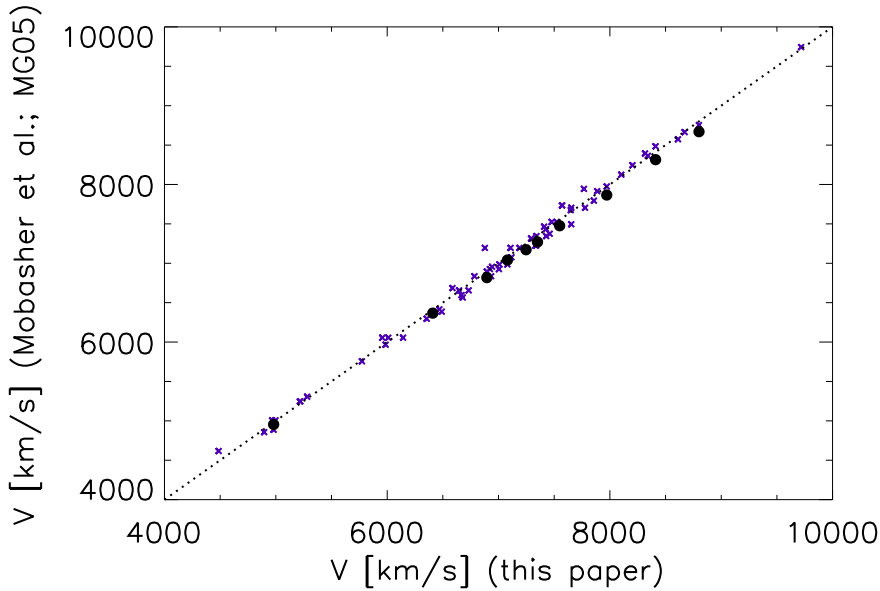}}
\end{center}
\caption{\label{velcomparison}
Velocity from Mob01 (crosses) and from MG05 (filled circles)
plotted against velocity from this paper. The straight line shows equality, indicating
that the MG05 values are systematically lower.
}
\end{figure}

We compare our radial velocities with those derived from our WYFFOS observations of a larger 
spectroscopic sample (Mob01). All of our galaxies already have velocities 
from that paper, and in Figure~\ref{velcomparison} we compare the velocities. The mean 
difference between the two datasets is 2.7~km~s$^{-1}$, in the sense that the WYFFOS 
velocities are higher, and the scatter about the linear relation is 81~km~s$^{-1}$. Most of 
the scatter can be attributed to errors on the WYFFOS velocities, which were obtained at lower 
resolution. Ten of our galaxies are common with the sample of MG05. These velocities are also 
compared in Figure~\ref{velcomparison}. The mean difference in this case is 75~km~s$^{-1}$, in 
the sense that our velocities are higher, with a scatter of 32~km~s$^{-1}$.  We conclude that 
there is a systematic offset in the velocity system of the MG05 data, of about 75~km~s$^{-1}$, 
but that the random error in each dataset is of the order 32/$\sqrt{2}\sim$22~km~s$^{-1}$. The 
random error is our WYFFOS dataset is probably twice this.

\begin{figure}
\begin{center}
{\includegraphics[width=11cm]{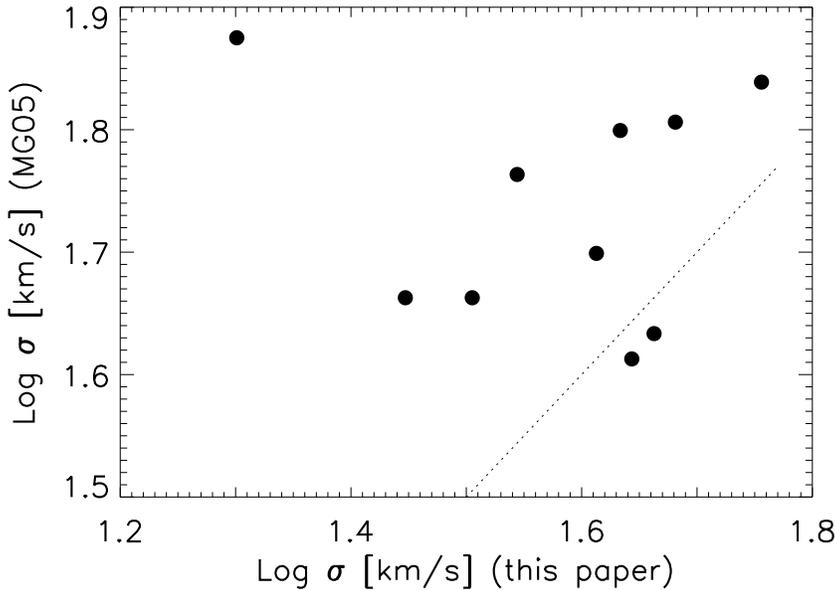}}
\end{center}
\caption{\label{dispcomparison}
Velocity dispersions from MG05 plotted against those from this paper.
The straight line shows the locus of equal velocities; MG05 velocity dispersions are systematically
higher.}
\end{figure}

In Figure~\ref{dispcomparison} we compare the velocity dispersions with those of
MG05 for the ten galaxies in common. There is considerable scatter about the 45
degree line in this plot; however this plot only covers a small range in
$\log{\sigma}$. We find that a formal fit gives an offset in $\log{\sigma}$ of
0.157, in the sense that MG05 dispersions are higher, and an rms of 0.171 about
this value. However a substantial part of this is due to a single galaxy, GMP3018,
for which we measure a dispersion of 20 km/s, and MG05 measure 75 km/s. The cause of
this is unclear, it could be due to a misidentification in one of the two studies, or
it could be due to poor signal-to-noise (of the ten galaxies in common, GMP has the
lowest signal-to-noise in both studies).

\begin{figure}
\begin{center}
\rotatebox{0}{\includegraphics[width=7.5cm]{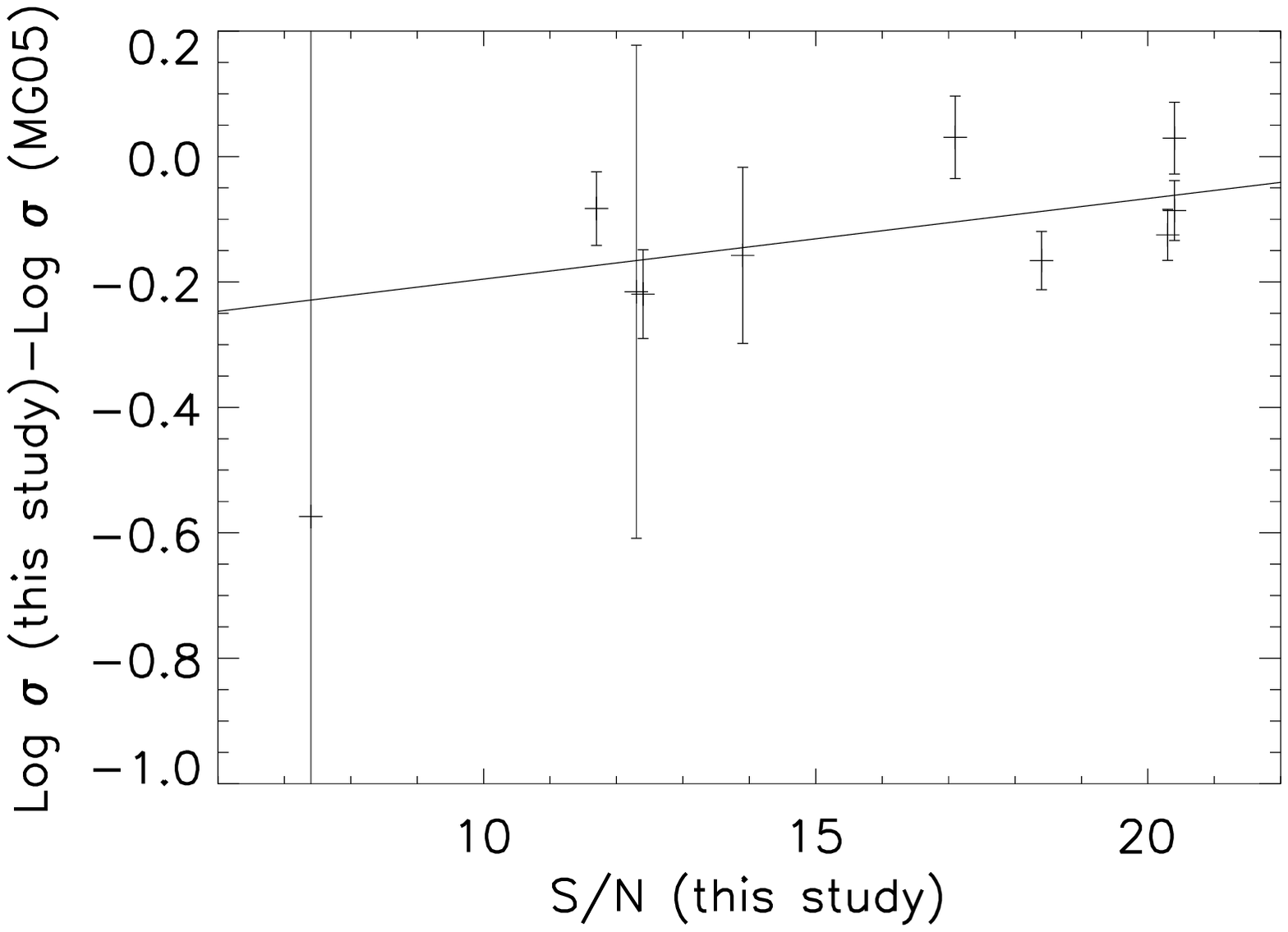}}
\rotatebox{0}{\includegraphics[width=7.5cm]{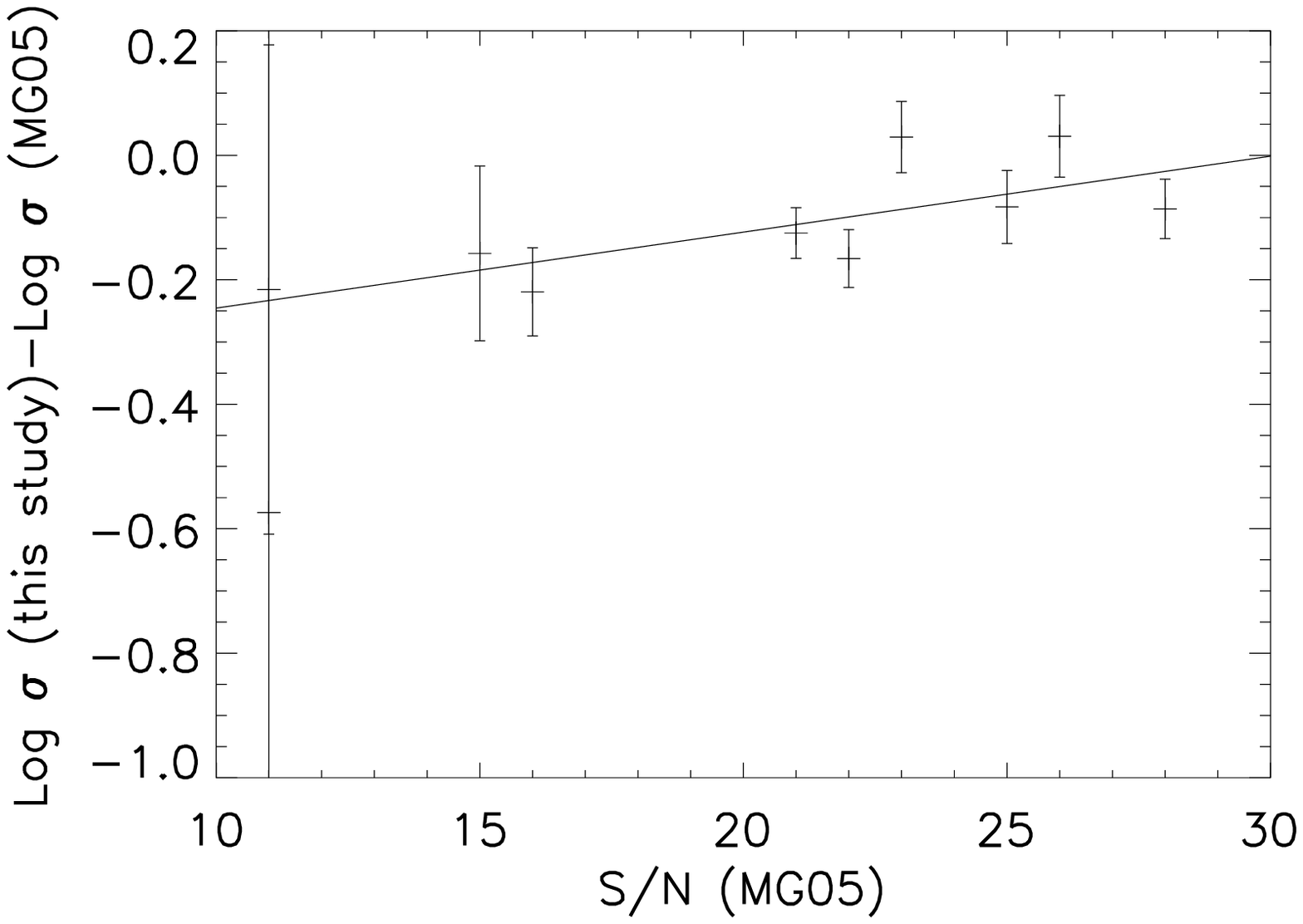}}
\end{center}
\caption{\label{dispcomparison_SN} Offset in log~$\sigma$ between our study and MG05,
plotted against the signal to noise in our study (left panel) and in MG05 (right panel).
The correlation coefficients are 0.52 and 0.74 respectively. The fits shown and discussed
in the text omit the most discrepant galaxy, GMP3018.}
\end{figure}

At the suggestion of the referee, we have investigated whether the offset is systematic, as 
might be caused by a calibration error in one study or the other, or is due to a signal-to-noise 
problem. In Figure~\ref{dispcomparison_SN} we plot the differences in $\log{\sigma}$ against S/N 
from each study. The most likely cause of the differences would appear to be effects due to the 
poor signal-to-noise in the fainter galaxies. Given that our spectral resolution is a factor two 
higher than MG05, and given also that the strongest correlation is with their S/N, it is likely 
that our dispersions are reliable to lower values. However, even with our resolution, 
dispersions below 30~km~s$^{-1}$ are likely to be systematically too high.

In figure~\ref{dispcomparison_SN} we show also linear least-squares fits as a function of
signal-to-noise. The linear fits are:
$$\Delta(\log\sigma) = -0.32 (\pm 0.13) + 0.013 (\pm 0.008) * S/N ({\rm this~study})$$
$$\Delta(\log\sigma) = -0.37 (\pm 0.09) + 0.012 (\pm 0.004) * S/N ({\rm MG05})$$
In section~\ref{trends} we will use the second of these two relations to transform the MG05
dispersions onto our system. We estimate that this fit only applies for S/N values up to 
30, above which there is sufficient agreement between the two datasets. Due to the large 
measurement uncertainties, we do not fit higher order functions to the data.

To ensure that our lower dispersion measurements are not strictly the result of the method 
of computing velocity dispersions, we have performed an additional analysis with the 
Penalized Pixel-fitting method (pPXF; Cappellari \& Emsellem 2004). This approach 
reconstructs the line-of-sight velocity distribution via a parametric fit to the galaxy 
spectra using large numbers of stellar templates. For high signal-to-noise data, it permits 
the extraction of higher order moments of the velocity distribution. However, for the lower 
S/N values of our spectra, we do not expect to accurately measure departures from a gaussian 
profile. Hence, we employ pPXF as a check on the dispersions for the subset of ten galaxies in 
common between our data and those of MG05, but retain FXCOR as our method of choice for the 
overall results.

For each galaxy in the comparison, we have computed velocity dispersions with both FXCOR 
and pPXF using several different sets of stellar template spectra. In addition to the five 
templates obtained with our Hydra setup, we have incorporated large numbers of spectra from 
the Indo-U.S. Library of Coud\'{e} Feed Stellar Spectra (Valdes {\sl et al.} 2004 ). The 
pPXF program efficiently computes velocity dispersions for large sets of templates, thereby 
allowing us to eliminate the effects of template mismatch. Among the template sets used to 
measure dispersions were five Coud\'{e} Feed Library (CF) templates with spectral types 
matched to those of our Hydra spectra, 60 CF templates with spectral types ranging from A 
through M, 40 CF templates with spectral types from G to M, and two CF spectra for stars in 
our Hydra sample (HD 62509 and HD 65583). To compare the different realizations of 
dispersions, we have degraded the CF spectra to the resolution of Hydra and computed the 
average offset between measurements of the same galaxies. To determine the significance of 
these differences, we apply the Kolmogorov-Smirnov (KS) test to different pairs of 
dispersion measurements on all ten galaxies. Our results indicate that FXCOR and pPXF 
return galaxy velocity dispersions that agree to within 5\% ($<2$ km~s$^{-1}$ offset, of 
the order the uncertainties), or a significance of 98\% when the same template set is used. 
This confirms that the measurements are not dependent on the computational method. We also 
find that regardless of templates or method used, dispersions measured for our ten galaxies 
in common with MG05 are systematically offset by $\sim 40\%$, at a significance of 90--95\% 
(i.e., only a 5--10\% chance that these differences are due to random variation). We 
conclude that the difference between our measurements and those of MG05 is real, at least 
at the low dispersions of the 10 galaxies in question. 

In addition, we have used the results of our tests with pPXF to assess the dependence of the 
velocity dispersions on the number of templates and their spectral types. The majority of 
these experiments produced results in agreement with our original measurements using FXCOR 
and Hydra templates. However, in a few cases the agreement is only marginal (67\% level). 
Surprisingly, the most discrepant velocity dispersions resulted from a trial of five CF 
templates with spectral types matching those of the five Hydra templates. The two sets of 
templates produced dispersions differing by an average of 5 km/s, a disagreement at the 31\% 
level). This cannot be due to template mismatch, and we attribute the disparity to 
differences in instrumental setup and resolution (despite the fact that we have degraded the 
CF spectra to match ours). We also note that in the cases where CF library templates yield 
significantly different velocity dispersions from those of the Hydra templates, the 
measurements are always {\em lower}. Hence, this does not explain the even larger discrepancy 
with the MG05 results. Because there appears to be an unknown source of $\sim 3$~km~s$^{-1}$ 
variation in the results that is not accounted for by template mismatch, we incorporate it 
as an additional uncertainty to our velocity dispersions. Errors listed in 
Table~\ref{resultstable} are thus the formal uncertainties calculated in \S3.3, plus a 
template mismatch error given by the average difference between dispersions from different 
Hydra templates, plus the $3$~km~s$^{-1}$ additional uncertainty, all added in quadrature.

\subsection{Trends with velocity dispersion}
\label{trends}

\begin{figure}
\begin{center}
{\includegraphics[width=12cm]{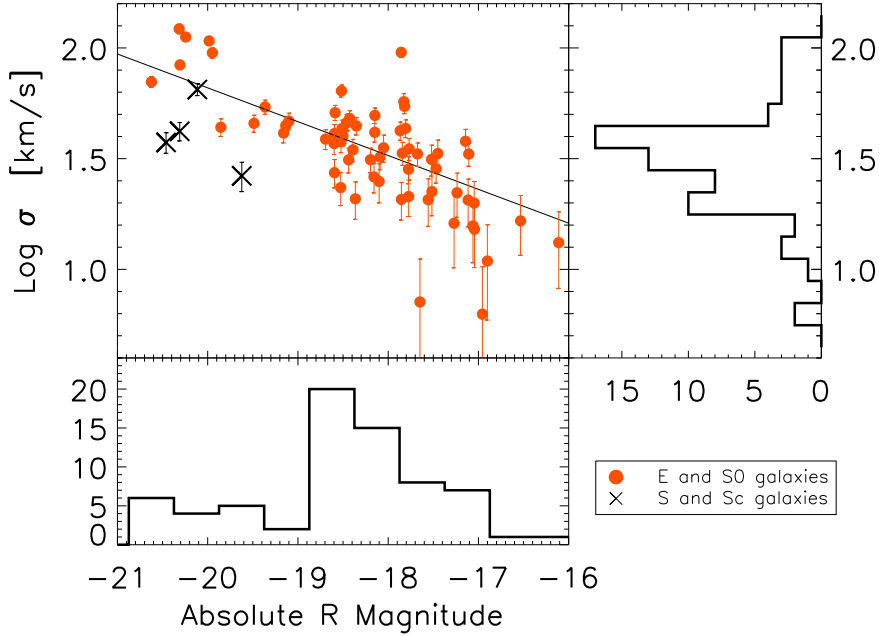}}
\end{center}
\caption{\label{Lsigmaplot}
Relation between $\log{\sigma}$ and absolute magnitude for our sample. In this plot, 
the different symbols represent different morphological classifications: filled circles
are ellipticals or S0 galaxies, and crosses are galaxies classified as spirals or 
possible spirals. The fit shown by the straight line is given by $\log{\sigma} = 
-2.1{\pm}0.4-(0.20{\pm}0.02) M_R,$ and the points representing spirals have been excluded.}  
\end{figure}

To assess our data in the context of the well-known properties of giant ellipticals, we have 
plotted a number of photometric parameters for galaxies from Mob01 
against the velocity dispersions. The velocity dispersion values for our galaxies are almost 
exclusively less than 100~km~s$^{-1}$, while the vast majority of those measured for galaxies 
traditionally considered ``giants'' are greater than 100~km~s$^{-1}$, and often 
200~km~s$^{-1}$.  In Figure~\ref{Lsigmaplot} we plot the relation between log~$\sigma$ and 
absolute magnitude in the {\em R} band. The distribution of parameters can be seen in the 
accompanying histograms. The absolute magnitudes in this fit were derived from our R band 
apparent magnitude within three Kron radii (Mob01), again assuming a 
distance modulus of 35.1 magnitudes, and an extinction in the R band of 0.03 magnitudes 
(Bernstein {\sl et al.} 1995). Different symbols in this plot represent different 
morphological types (from Table~\ref{resultstable}).

Our sample contains four galaxies which according to our visual classification are, or 
might possibly be, spirals (crosses in Figure~\ref{Lsigmaplot}). In calculating our 
best-fitting regression line, we exclude these galaxies, and find a relation $\log{\sigma} = 
-2.1{\pm}0.4-(0.20{\pm}0.02) M_R$. In this and all other regression fits we minimise residuals 
in $\log{\sigma}$, as the errors in this quantity are far larger than those on the photometric 
parameters. This gives an $L-\sigma$ relation of the form $L_R \propto \sigma^{2.0{\pm}0.2}$, 
which is substantially flatter than the Faber-Jackson relation for giant ellipticals ($L 
\propto \sigma^4$, Faber \& Jackson 1976) but consistent with results presented by MG05 and 
previous authors.

\begin{figure}
\begin{center}
{\includegraphics[width=12cm]{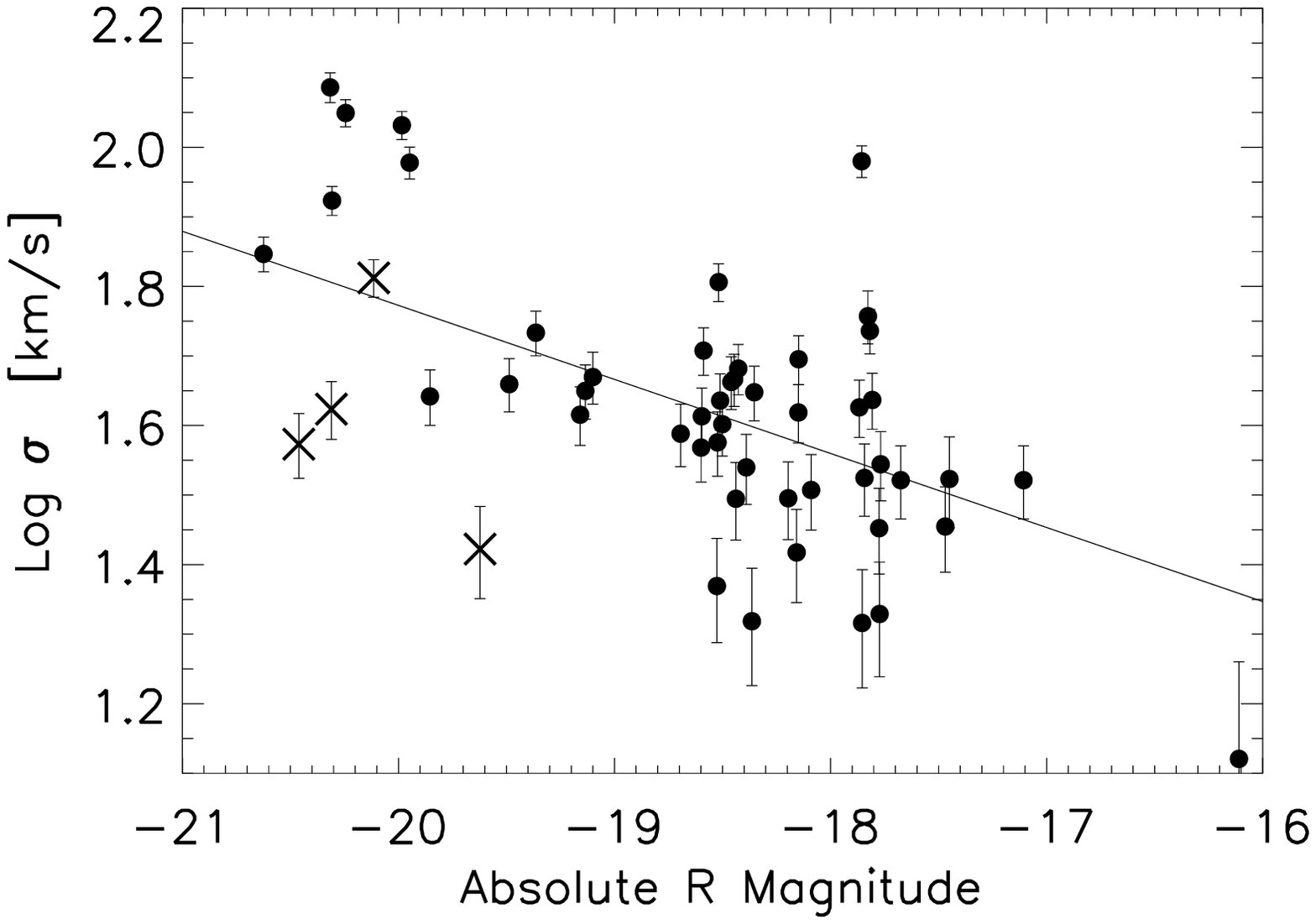}}
\end{center}
\caption{\label{fig:HighSNonly} Relation between $\log{\sigma}$ and absolute magnitude for 
the E and S0 galaxies with $S/N > 10$. Symbols are the same as in Fig.\ 7. The fit to the high 
$S/N$ galaxies is given by $\log{\sigma} = -1.2{\pm}0.4-(0.16{\pm}0.04) M_R$.} 
\end{figure}

To investigate whether the inclusion of galaxies with low signal-to-noise spectra might bias the 
results, we have repeated the analysis, excluding those galaxies with spectra with $S/N < 
10$. In figure \ref{fig:HighSNonly} we plot the remaining galaxies, and the fit to the slope of the 
$L-\sigma$ relation. We find $\log{\sigma} = -1.2{\pm}0.4-(0.16{\pm}0.04) M_R$, giving
$L_R \propto \sigma^{2.5{\pm}0.8}$, so the slope does not differ significantly from that for 
the whole sample. 

\begin{figure}
\begin{center}
{\includegraphics[width=12cm]{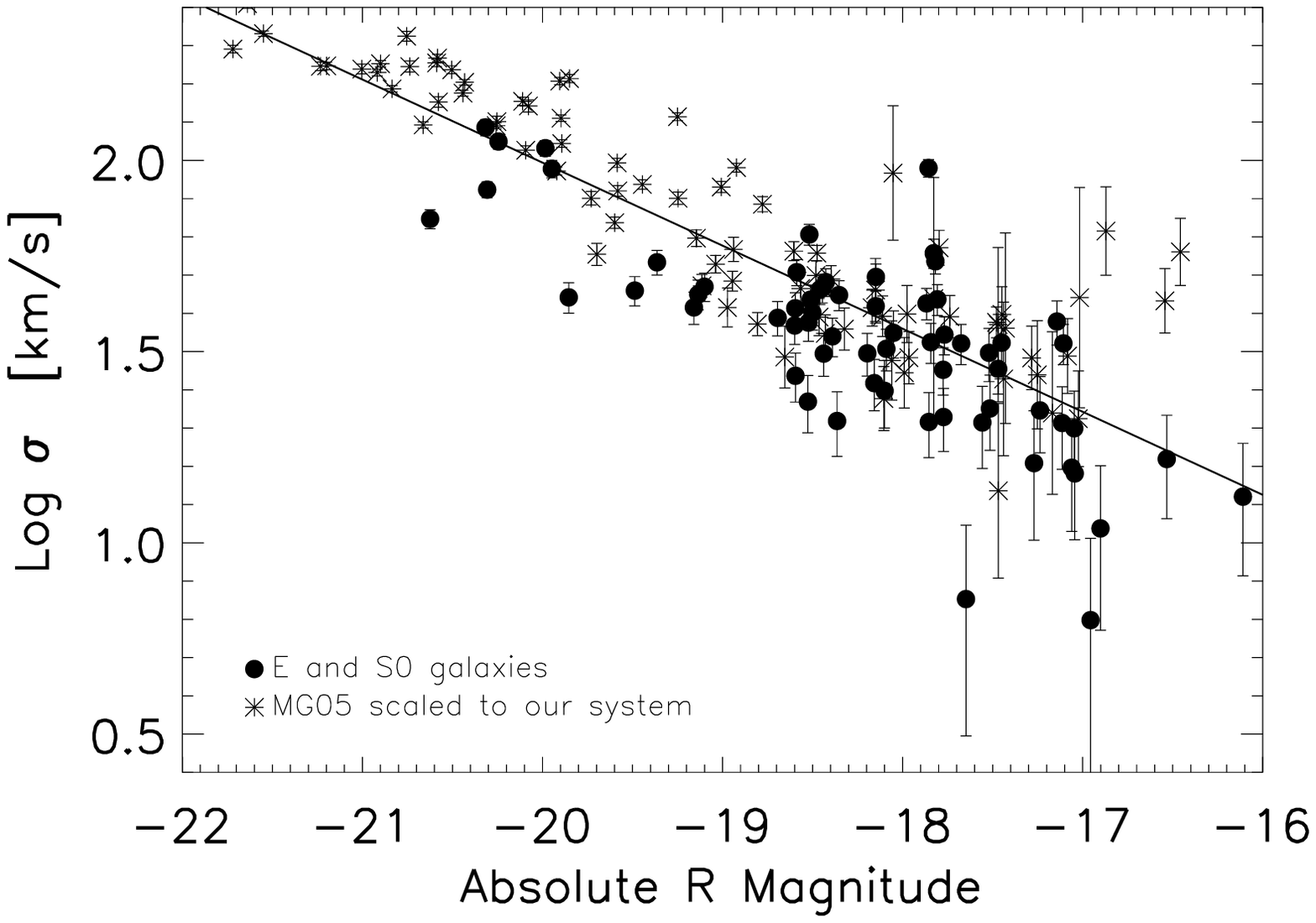}}
\end{center}
\caption{\label{Fig:plotwithMG05} Relation between $\log{\sigma}$ and absolute magnitude for 
the E and S0 galaxies in our sample and for the sample of MG05, adjusted to
our velocity dispersion system as described in the text (asterisks).
The fit shown by the straight line is to both samples, with galaxies in both samples
included twice, and is given by $\log{\sigma} = -2.35{\pm}0.21-(0.217{\pm}0.011) M_R$.} 
\end{figure}

To increase the size of our sample at the brighter end of this correlation, for those
galaxies in common between MG05 and Mob01 we have moved the MG05
dispersions onto our system, using the second of the two linear correlations presented
in Section~\ref{comparisons}. In figure~\ref{Fig:plotwithMG05} we plot our dispersions   
and the transformed MG05 values, together with a fit showing an $L-\sigma$ relation of  
the form $L_R \propto \sigma^{1.84{\pm}0.10}$, entirely consistent with that defined by
our sample alone.

\begin{figure}
\begin{center}
{\includegraphics[width=12cm]{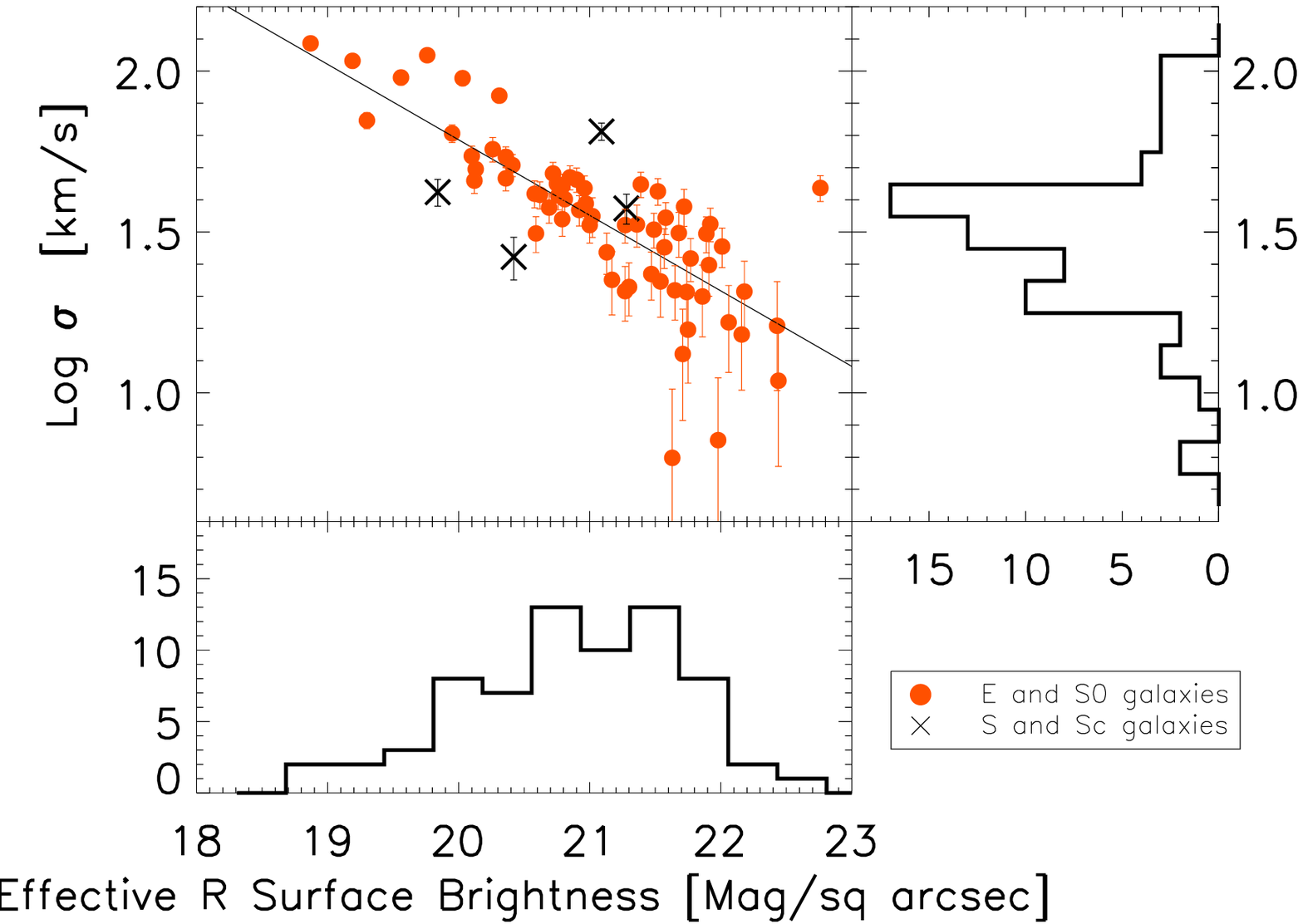}}
\end{center}
\caption{\label{SBplot}
Log $\sigma$ against effective {\em R} band surface brightness. The symbols have the same 
meaning as in Figure~\ref{Lsigmaplot}. The trend is given by $\log{\sigma} = 
6.7{\pm}0.5-(0.25{\pm}0.02) \mu_{\rm eff}$. Again, the fit has been calculated omitting the 
spirals. 
}  
\end{figure}

In Figure~\ref{SBplot} we plot log~$\sigma$ against the effective surface brightness from 
Mob01. The best-fitting linear regression line is $\log{\sigma} = 
6.7{\pm}0.5-(0.25{\pm}0.02) \mu_{\rm eff}$, where $\mu_{\rm eff}$ is the average {\em R} band 
surface brightness within the effective radius, in magnitudes arcsec$^{-2}$. This corresponds 
to a relation of the form $I_m \propto \sigma^{1.6{\pm}0.2}$, where $I_m$ is effective 
surface brightness in flux units, in the R band. The relationship between surface brightness 
and velocity dispersion also differs from that of the giant ellipticals. Using the result of 
$L \propto I_m^{-1.5}$ from Binggeli, Sandage \& Tarenghi (1984) with the Faber Jackson relation, we 
expect $I_m \propto \sigma^{-2.5}$ for giant elliptical galaxies. Our value for the 
lower-luminosity galaxies in this sample is thus quite different.

Another trend to explore is that of velocity dispersion and effective radius. The latter is 
directly related to luminosity and effective surface brightness via $L=2\pi IR_e^2.$ Using our 
measured relationships between $\sigma$, luminosity and surface brightness we derive $R_e 
\propto \sigma^{0.19{\pm}0.14}$. This is just consistent with the relation that would be 
derived using the established $L-R_e$ trend for dwarf ellipticals (Binggeli {\sl et al.} 1984), 
which is $L \propto R_e^{4}$ and the measured $L-\sigma$ relation from MG05 and our work, which 
would together give $R_e \propto \sigma^{0.5}$. It is however inconsistent with the relation 
for giants, derived from Fish's (1964) law ($L \propto R_e^{1.2}$) and the Faber-Jackson 
relation, which together imply $R_e \propto \sigma^{3.3}$.

Since the mass-to-light ratio, $M/L$, is an important indicator of galaxy properties, we have 
computed its variation with $\sigma$ for our dwarf sample. Assuming the galaxies we have 
observed form a homologous sequence, we can compute the expected $M/L$ as a function of 
$\sigma$ by applying the virial theorem (e.g., Mobasher {\sl et al.} 1999; D'Onofrio {\sl et 
al.} 2006). This results in ${M/L} \propto {{\sigma^2}\over {I_m R}}$ (e.g.  Richstone \& 
Tremaine 1986).  Regarding the radius as an independent parameter, we express it in terms of 
surface brightness and luminosity to obtain ${M/L} \propto {{\sigma^2} \over \sqrt{I_m L}}$.  
Substituting our observed dependence of $L$ and $I_m$ upon $\sigma$, we obtain ${M/L} \propto 
{\sigma^{0.19{\pm}0.14}}.$

\subsubsection{Trends with Radius within the cluster}

The mechanisms proposed to explain the differences in scaling laws between giants and dwarfs 
(Section~\ref{introduction}) may be affected by environment. In particular, winds may be 
constrained and ram-pressure stripping enhanced by the hot gas density in cluster cores, and 
the importance of harassment and tidal dwarf formation depends upon both the galaxy density 
and velocity dispersion. Furthermore, Smith {\sl et al.} (2008), working in precisely the same 
regions of the Coma cluster, found a significant environmental dependence of the stellar 
population properties in dwarf galaxies, and thus evidence of environmental influence on the 
star formation history. As our sample includes galaxies in the infall region of the cluster, we 
can search for a possible environmental dependence of the $L-\sigma$ relation, as a counterpart 
to the environmental dependence of the properties of the stellar populations. To undertake this 
comparison we divide our sample into the Coma1 and Coma3 regions of Kom02, 
which separates the sample at a clustocentric radius of 0.49 degrees. However, our 
sample has a different luminosity distribution within these two regions; the outer Coma3 sample 
contains many more galaxies with $M_R < -19$, so we limit this analysis to galaxies with $M_R > 
-19$.

\begin{figure}
\begin{center}
{\includegraphics[width=12cm]{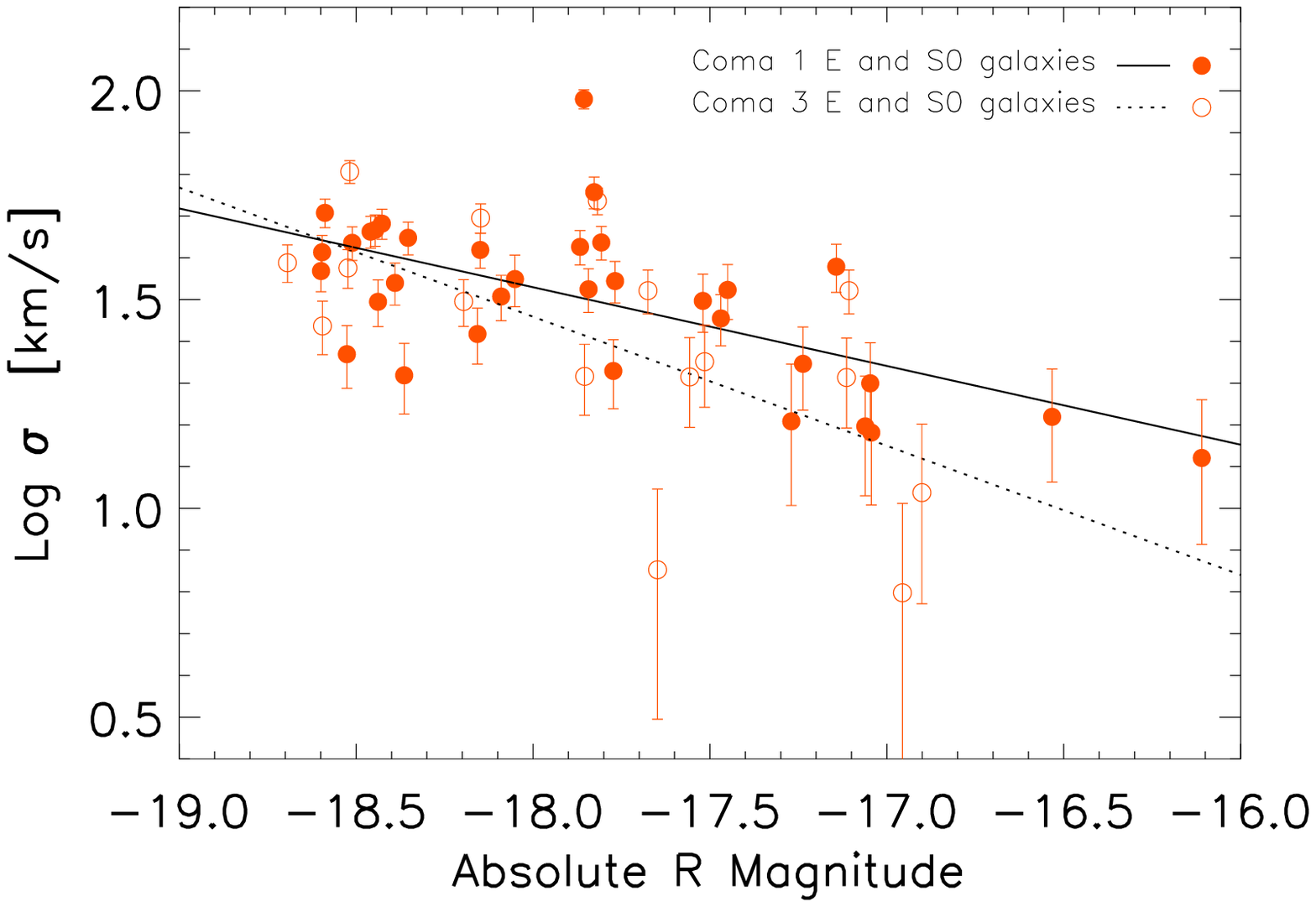}}
\end{center}
\caption{\label{Fig:Faintplot} Relation between $\log{\sigma}$ and absolute magnitude for 
the E and S0 galaxies with $M_R > -19$ in the Coma1 field inside 0.49 degrees radius (filled circles),
and in the Coma3 field outside 0.49 degrees radius (open circles). 
The solid line shows the fit to the inner field, and is $\log{\sigma} = 
-1.87{\pm}0.73-(0.19{\pm}0.04) M_R$, while the dashed line is the fit to
the outer field, $\log{\sigma} = -4.11{\pm}1.80-(0.31{\pm}0.10) M_R$.}

\end{figure}

In Figure \ref{Fig:Faintplot} we show the $L-\sigma$ relation for our inner and outer fields.
The slope of the fit is shallower in the outer field ($L_R \propto \sigma^{1.29{\pm}0.29}$ as
opposed to $L_R \propto \sigma^{2.21{\pm}0.40}$). Despite the marginal significance of the difference 
we find, this possible environmental effect merits further investigation by increasing the sample, 
and thus lowering the error bar, in the outer region.

\section{Discussion}
\label{discussion}

Our results confirm that galaxies in the dwarf regime do not follow the standard empirical 
relations established for giant ellipticals. Of primary interest is the fact that the $L-\sigma$ 
relation for dwarfs is substantially shallower than the classical Faber-Jackson relation.  This 
was initially suggested by Davies {\sl et al.} (1983), whose kinematic study of 11 faint 
elliptical galaxies with $M_R > -20.8$ led to $L\propto\sigma^{2.4\pm 0.9}$. Many independent 
studies have found a relation between $L\propto\sigma^{2.0}$ and $L\propto\sigma^{2.55}$ in the R 
or B band (Held {\sl et al.} 1992; De Rijcke {\sl et al.} 2005; MG05). Peterson \& Caldwell 
(1993), on the other hand, found a much steeper $L_V\propto\sigma^{5.6}$ by including the local 
group dwarf spheroidal galaxies. Our study indicates that in an intermediate luminosity range, 
$-19 < M_R < -16$, galaxies continue to follow the shallower relation. In addition, the observed 
relation between surface brightness and velocity dispersion is different from that of giant 
ellipticals but consistent with that implied by the position of dEs in Figure~1c of De Rijcke {\sl 
et al.} (2005).

To confirm that the derived scalings for dwarf galaxies are substantially different from those 
of the giants, we must assess potential sources of bias.  Surface brightness selection and 
morphological misclassification of low surface brightness galaxies can bias the sample-- although 
elliptical and lenticular galaxies follow the same $L-\sigma$ relations (J{\o}rgensen \& Franx 
1994), spiral galaxies do not. Performing a linear regression on {\em all} of the data 
plotted in Figure~\ref{Lsigmaplot}, we find that the effect of contamination by spirals on the 
$L-\sigma$ relation is to increase the index; thus inclusion of late-type galaxies does not 
explain the difference from the Faber-Jackson relation for giants. Likewise, the difficulty in 
measuring velocity dispersion for the low surface brightness galaxies could produce 
erroneously high $\sigma$ values, but would also have the effect of raising the index in the 
$L-\sigma$ relation. The same reasoning holds for the run of surface brightness versus 
velocity dispersion; the index of the $I_m$-$\sigma$ relation would increase due to spiral 
contamination, and the implied $L-R_e$ relation would only change slightly.

Finally, we address the possibility that the luminosity and velocity dispersion values are 
systematically affected by the way we have defined these parameters. Luminosities have been 
derived from Kron magnitudes, which have been shown to agree with standard fixed-aperture 
magnitudes to within 2\% (Mob01). If this is a cause of measurement 
bias, such a percent-level effect would be well within the scatter of the data. Velocity 
dispersions must also be carefully defined, as some studies report central values, while 
others provide averages over a particular galactic radius. Fortunately, radius-dependent 
studies of velocity dispersion have shown that it varies only on the order of 10\% out to 
large radii (e.g., Bender \& Nieto 1990). Our own velocity dispersions have been measured 
over a 3\arcsec\ diameter, equivalent to 1.53 kpc. As discussed in \S~3.2, we believe 
this is preferable to obtaining a truly central measurement, to sample kinematics that are 
not dominated by the nucleus. In addition, a number of studies on giants to which we compare 
our data have also used normalized apertures, as suggested by J{\o}rgensen, Franx, \& 
Kj{\ae}rgaard (1996). With this approach, J{\o}rgensen (1999) obtained velocity dispersions 
for Coma giants within a 3.4\arcsec diameter, corresponding to 1.73 kpc. The similar 
aperture size employed in the present study allows for a direct comparison with these data. 
Thus we are confident that the differences between the relations shown in 
Figures~\ref{Lsigmaplot} and ~\ref{SBplot}, and the equivalent relations for brighter 
galaxies, reflect the intrinsic properties of the samples, and are not due to biases in the 
sample selection or the methods employed to make the measurements.

It is unclear whether there is a true dichotomy between giants and dwarfs, or simply a 
continuous set of relations with gradually changing slopes. 
Nevertheless, we can explore the possible reasons for the observed dwarf galaxy scalings by 
comparing them with theoretical predictions based on the physical processes thought to operate 
in these galaxies. Although rotation was initially suggested as a cause for the differing 
kinematic parameters, and specifically the flattening of dwarf galaxies, recent work by Simien 
\& Prugniel (2002), Geha {\sl et al.} (2003), and MG05 has cast doubt on this idea. The 
anisotropic velocity dispersions observed in dwarfs have been also accounted for by invoking 
galactic winds spawned by supernovae, as suggested by Bender \& Nieto (1990). Winds are 
thought to dominate in objects with velocity dispersions under 60~km~s$^{-1}$ 
(Schaeffer \& Silk 1988). The processes of gas loss, galaxy inflation and associated 
re-adjustment to a new virial equilibrium provide an attractive way to explain differences 
between the scaling relations of dwarf and giant ellipticals. We consider this theory more 
promising than other mechanisms such as dissipation, merging, or tidal stripping of disk 
galaxies, since these do not appear capable of moving galaxies in the direction of the region 
that dwarfs occupy in surface brightness and velocity dispersion parameter space (e.g., Bender 
\& Nieto 1990; Bender {\sl et al.} 1992; De Rijcke {\sl et al.} 2005). Hence, we use the 
predictions of galactic wind models from DS86 and YA87 to test this scenario against the trends 
seen in our data.

The models of DS86 describe gas loss from self-gravitating systems with
similar age and initial stellar mass function. If the galaxy's dynamics are dominated by
a dark matter halo, they predict $L\propto\sigma^{5.26}$ and $M/L\propto L^{-0.37},$
leading to $M/L\propto M^{-0.59}\propto\sigma^{-1.95}.$ If instead the galaxies
contain only baryonic matter and have roughly constant $M/L$ (because of the similar
stellar content), they find that after gas loss, dwarfs should adhere to
$L\propto\sigma^{2.7}.$ YA87 provide a finer level of distinction with
two different rates of gas expulsion: slow and rapid. Such models specify a fraction of
the galaxy's gas mass to be blown out, the time-scale on which it happens, and the new
galaxy structure from the resulting virial equilibrium (although in the case of rapid gas
removal, equilibrium is not necessarily reached). For dwarf galaxies that have undergone
a complete course of slow, adiabatic gas removal, they obtain $L\propto\sigma^{2.5}$ and
mass-to-light ratios that scale inversely with luminosity. However, it should be noted
that these scaling relations only apply to the end stage of gas removal; if not all dwarf
galaxies have reached this point, then a substantial scatter in their properties might be
expected, as suggested by Figure~9 of YA87. De Rijcke {\sl et al.} (2005) overlay a number of 
late-type galaxies on these galactic wind models, showing a consistency between the positions of 
dwarf ellipticals, and models including gas removal.

Our data indicate a slow positive trend of $M/L$ with velocity dispersion, of the form 
$M/L\propto\sigma^{0.19}$. Whilst this is in broad agreement with the galactic wind models 
discussed above, it is very different from results from giant galaxies. For giants, the exponents 
in the range 0.66 to 1.4 have been found in the optical (J{\o}rgensen {\sl et al.} 1996,1998;  
Guzm\'{a}n, Lucey \& Bower 1993) and the near infra-red (Mobasher {\sl et al.} 1999). It is also 
different from the trend found for the much lower luminosity local group dwarfs. There Peterson \& 
Caldwell (1993) find $M/L\propto L^{-0.40\pm0.06}$ which, together with their measured relation 
$L\propto\sigma^{5.6\pm 0.9}$, yields $M/L\propto\sigma^{-2.2\pm 0.5}.$

Several recent studies have suggested a large variation of slope in the $M/L-\sigma$ relation 
over many orders of magnitude in galaxy luminosity. From data on a number of clusters, 
Zaritsky, Gonzalez \& Zabludoff (2006) have revealed a dependence that is roughly parabolic and 
not well represented by a power law. Their trend of $M/L$ on $\sigma$ becomes flat near 
$\log{\sigma}=1.7$ and is very consistent with our result of ${M/L} \propto \sigma^{0.19}$ 
obtained for galaxies with dispersions in that regime. Desroches {\sl et al.} (2007) also find 
significant curvature in the $M/L-\sigma$ relation among a large set of galaxies from the Sloan 
Digital Sky Survey, although their sample primarily encompasses objects with significantly 
larger luminosities and velocity dispersions than our dwarf data. Cappellari {\sl et al.} 
(2006) have calculated $M/L$ values based on dynamical models of the SAURON galaxy sample 
(Bacon {\sl et al.} 2001) and conclude that non-homology effects play only a small role in 
setting the fundamental scaling relations among late-type galaxies; stellar population effects 
and dark matter properties must instead account for changing mass-to-light ratios. 


\section{Conclusions}
\label{conclusions}

We have presented velocity dispersions for a sample of 69 dwarf and giant galaxies in the Coma 
cluster, of which 62 are either elliptical or S0 galaxies in their morphology.  We find that 
the fundamental parameters of low-luminosity ellipticals vary fairly tightly with the velocity 
dispersion. The relationship between luminosity and velocity dispersion at the boundary of the 
dwarf and giant regime is clearly different from the classical Faber-Jackson relation for 
giant elliptical galaxies.  The correlation between surface brightness and velocity dispersion 
in our sample is also very tight and can be explained by observational error in the velocity 
dispersions alone. There is evidence that the mass-to-light ratio varies systematically along 
this correlation, with the higher surface brightness galaxies having higher ($M/L$). We have 
investigated whether gas removal by galactic winds can explain these results and find that it 
provides a satisfactory origin for the $L-\sigma$ relation and a possible basis for the trend 
in mass-to-light ratio.

\section*{Acknowledgments}

The WIYN Observatory is a joint facility of the University of Wisconsin-Madison, 
Indiana University, Yale University, and the National Optical Astronomy Observatories.
A.M.C. acknowledges the AAO Student Fellowship Program for support of this work.
We would like to thank Pat Knezek and Megan Sosey for help with the observations,
as well as WIYN Hydra Instrument Scientist Diane Harmer for assistance with 
observing preparations. We thank the anonymous referee for helpful comments.

\clearpage

\appendix

\section{Velocity and velocity dispersion data}

{\scriptsize

\begin{longtable}[tc]{lllllllllllll}
\caption{
\label{resultstable}
Velocities and velocity dispersions of Coma dwarfs}

\endhead

\endfoot

\hline 
GMP	&	Komiyama	&	Type	&	RA	&	Dec 		&	Kron R 	&	B-R	&$\mu_{\rm R,eff}$&	cz 	&	Error	&	 $\sigma$	&	Error	&	 S/N \\
ID&ID&&(J2000)&(J2000)&mag.&mag.&mag/(\arcsec)$^2$&km s$^{-1}$&km s$^{-1}$ &km s$^{-1}$&km 
s$^{-1}$ & (per \AA) \\ \hline
4644	&	48397	&	dE	&	12 58 34.1	&	26 53 60	&	18.17	&	1.37	&	21.63	&	8103	&	5	&	6	&	4	&	7.0\\
5361	&	50320	&	E	&	12 56 36.0	&	26 54 18	&	16.44	&	1.42	&	20.97	&	7886	&	4	&	39	&	4	&	13.7\\
4518	&	55692	&	E	&	12 58 03.2	&	26 54 58	&	14.51	&	1.41	&	19.30	&	8203	&	6	&	70	&	4	&	16.6\\
4340	&	53772	&	E/S0	&	12 58 20.3	&	26 55 14	&	16.93	&	1.24	&	20.59	&	6934	&	3	&	31	&	3	&	20.9\\
4418	&	59497	&	dE	&	12 58 13.0	&	26 56 59	&	18.02	&	1.47	&	21.74	&	6732	&	6	&	21	&	5	&	7.0\\
4980	&	65832	&	E/S0	&	12 57 18.6	&	26 58 47	&	17.46	&	1.42	&	21.27	&	7293	&	4	&	34	&	4	&	12.0\\
5395	&	83625	&	S0	&	12 56 32.0	&	27 03 21	&	14.81	&	1.37	&	18.87	&	6143	&	8	&	122	&	5	&	28.6\\
5259	&	82181	&	dS0	&	12 56 47.1	&	27 03 25	&	17.57	&	1.22	&	22.18	&	6490	&	8	&	21	&	5    	&	6.6\\
5097	&	87820	&	dE	&	12 57 05.8	&	27 05 21	&	18.23	&	1.50	&	22.44	&	7478	&	5	&	12	&	4	&    	7.4\\
5032	&	92047	&	S0	&	12 57 12.0	&	27 06 13	&	16.03	&	1.40	&	20.85	&	7331	&	4	&	47	&	3	&	17.7\\
4135	&	103965	&	SBc	&	12 58 37.3	&	27 10 35	&	14.82	&	1.06	&	19.84	&	7654	&	9	&	42	&	3	&	33.3\\
5365	&	115034	&	S0/a	&	12 56 34.6	&	27 13 40	&	15.28	&	1.25	&	20.79	&	7188	&	4	&	44	&	3	&	20.0\\
4591	&	113838	&	E	&	12 57 55.4	&	27 13 55	&	17.28	&	1.37	&	21.27	&	6469	&	4	&	21	&	4	&	12.0\\
5422	&	130251	&	Sc/SBc	&	12 56 28.6	&	27 17 29	&	14.67	&	1.15	&	21.28	&	7522	&	4	&	37	&	3	&	22.3\\
4351	&	131621	&	S0pec	&	12 58 18.7	&	27 18 38	&	15.77	&	1.18	&	20.36	&	7410	&	17	&	54	&	3	&	21.6\\
4565	&	138413	&	dE	&	12 57 58.0	&	27 21 03	&	18.02	&	1.46	&	21.00	&	8672	&	5	&	33	&	4	&	11.5\\
5364	&	144552	&	SB0	&	12 58 33.1	&	27 21 52	&	14.82	&	1.55	&	20.31	&	7009	&	5	&	84	&	4	&	21.6\\
5136	&	143923	&	SB0	&	12 57 01.7	&	27 22 20	&	15.18	&	1.52	&	20.03	&	7004	&	5	&	95	&	4	&	22.9\\
4375	&	145796	&	E	&	12 57 56.5	&	27 22 56	&	16.54	&	1.48	&	21.13	&	5221	&	3	&	27	&	4	&	14.7\\
4215	&	149036	&	dS0	&	12 58 31.7	&	27 23 42	&	17.48	&	1.29	&	21.98	&	7571	&	5	&	7	&	4	&	9.9\\
4479	&	153508	&	E/S0	&	12 58 06.1	&	27 25 08	&	16.00	&	1.43	&	20.75	&	5774	&	3	&	45	&	3	&	18.1\\
5250	&	154595	&	S0	&	12 56 47.8	&	27 25 16	&	15.64	&	1.40	&	20.12	&	7777	&	3	&	46	&	3	&	23.2\\
4430	&	156329	&	E/S0	&	12 58 20.5	&	27 25 46	&	16.61	&	1.319	&	20.69	&	7540	&	3	&	38	&	3	&	22.3\\
5526	&	161876	&	E/SB0	&	12 56 16.7	&	27 26 45	&	14.89	&	1.71	&	19.76	&	6404	&	8	&	112	&	5	&	25.1\\
5296	&	159473	&	S0	&	12 56 40.9	&	27 26 52	&	17.32	&	1.50	&	20.69	&	7338	&	3	&	43	&	3	&	17.2\\
4381	&	162274	&	dE/S0	&	12 58 15.3	&	27 27 53	&	17.62	&	1.52	&	21.17	&	7650	&	6	&	22	&	4	&	7.9\\
4956	&	164198	&	E	&	12 57 21.7	&	27 28 30	&	16.98	&	1.69	&	20.13	&	6942	&	4	&	50	&	3	&	15.9\\
4597	&	169748	&	SBc	&	12 57 54.4	&	27 29 26	&	15.02	&	1.36	&	21.09	&	4986	&	4	&	65	&	3	&	24.5\\
4522	&	167048	&	E	&	12 57 50.8	&	27 29 27	&	17.31	&	1.47	&	20.10	&	7341	&	4	&	55	&	3	&	16.4\\
5102	&	180920	&	S0	&	12 57 04.3	&	27 31 34	&	15.97	&	1.47	&	20.62	&	8341	&	3	&	42	&	3	&	18.3\\
4852	&	176486	&	E	&	12 57 30.6	&	27 32 35	&	16.61	&	1.60	&	19.95	&	7653	&	4	&	64	&	3	&	21.5\\
4117	&	180017	&	E	&	12 58 38.4	&	27 32 39	&	15.15	&	1.54	&	19.19	&	5986	&	7	&	108	&	5	&	26.1\\
5284	&	181166	&	S0	&	12 56 42.4	&	27 32 54	&	16.54	&	1.47	&	20.41	&	7571	&	3	&	51	&	3	&	20.9\\
3760	&	2623	&	dE	&	12 59 6.38	&	27 33 39	&	17.86	&	1.48	&	22.43	&	7767	&	6	&	16	&	5	&	6.0\\
3271	&	5443	&	S0/a	&	12 59 39.8	&	27 34 36	&	15.51	&	1.02	&	20.42	&	4997	&	5	&	26	&	3	&	32.1\\
3585	&	6728	&	S0	&	12 59 18.5	&	27 35 37	&	16.77	&	1.16	&	21.65	&	5216	&	5	&	21	&	4	&	19.9\\
3598	&	13606	&	dE	&	12 59 17.1	&	27 38 03	&	17.99	&	1.51	&	21.72	&	5282	&	5	&	38	&	4	&	9.8\\
2801	&	23231	&	dS0	&	13 00 17.4	&	27 42 41	&	17.89	&	1.57	&	21.54	&	7108	&	5	&	22	&	4	&	9.1\\
3586	&	23987	&	E	&	12 59 18.3	&	27 42 56	&	16.68	&	1.58	&	20.36	&	6679	&	3	&	46	&	3	&	20.7\\
3176	&	28211	&	S0	&	12 59 46.3	&	27 44 46	&	17.26	&	1.09	&	21.52	&	9718	&	7	&	42	&	4	&	14.6\\
4035	&	29543	&	E	&	12 58 45.5	&	27 45 14	&	16.97	&	1.50	&	21.77	&	6646	&	4	&	26	&	4	&	12.0\\
4150	&	31541	&	dE	&	12 58 35.4	&	27 46 30	&	18.60	&	1.41	&	22.06	&	5956	&	6	&	17	&	5	&	6.0\\
2478	&	39218	&	E/S0	&	13 00 45.4	&	27 50 08	&	16.53	&	1.59	&	20.76	&	8800	&	3	&	41	&	3	&	20.4\\
4175	&	39682	&	E/S0	&	12 58 33.8	&	27 50 12	&	17.08	&	1.36	&	21.02	&	4486	&	5	&	35	&	5	&	8.9\\
2753	&	41046	&	E/S0	&	13 00 20.2	&	27 50 36	&	16.53	&	1.60	&	20.92	&	7856	&	3	&	37	&	3	&	16.9\\
3473	&	42068	&	dE	&	12 59 26.4	&	27 51 25	&	17.68	&	1.54	&	21.36	&	4967	&	5	&	33	&	4	&	10.1\\
3645	&	46757	&	E	&	12 59 14.6	&	27 53 44	&	17.30	&	1.59	&	20.26	&	6409	&	6	&	57	&	5	&	11.7\\
2736	&	47098	&	E	&	13 00 21.6	&	27 53 55	&	16.74	&	1.49	&	20.79	&	4896	&	3	&	35	&	3	&	17.5\\
2376	&	47923	&	S0	&	13 00 55.9	&	27 53 55	&	16.69	&	1.60	&	21.89	&	6010	&	4	&	31	&	4	&	11.1\\
3080	&	49731	&	dE	&	12 59 55.7	&	27 55 03	&	18.07	&	1.72	&	21.75	&	6668	&	3	&	16	&	4	&	8.6\\
3511	&	50139	&	cE	&	12 59 23.4	&	27 55 10	&	17.27	&	1.66	&	19.56	&	6923	&	6	&	95	&	4	&	23.9\\
3376	&	50554	&	dE	&	12 59 32.0	&	27 55 15	&	17.66	&	1.49	&	22.01	&	7079	&	4	&	28	&	4	&	12.3\\
2692	&	52689	&	S0/a	&	13 00 24.8	&	27 55 36	&	16.78	&	1.62	&	21.39	&	7972	&	3	&	44	&	3	&	17.1\\
3565	&	57356	&	E/S0	&	12 59 19.7	&	27 58 24	&	17.04	&	1.53	&	21.49	&	7247	&	4	&	32	&	4	&	13.9\\
3602	&	58030	&	dE	&	12 59 16.7	&	27 58 57	&	19.02	&	1.58	&	21.71	&	6878	&	5	&	13	&	4	&	6.9\\
3018	&	59516	&	dE	&	13 00 01.0	&	27 59 30	&	18.08	&	1.62	&	21.86	&	7549	&	6	&	20	&	4	&	7.4\\
3166	&	59610	&	E	&	12 59 46.9	&	27 59 31	&	17.36	&	1.46	&	21.58	&	8410	&	4	&	35	&	4	&	12.4\\
3292	&	60593	&	S0/a	&	12 59 38.0	&	28 00 03	&	16.62	&	1.51	&	20.96	&	4980	&	3	&	43	&	3	&	18.4\\
3969	&	61500	&	E	&	12 58 50.8	&	28 00 25	&	17.36	&	1.55	&	21.30	&	7460	&	3	&	21	&	4	&	14.7\\
3681	&	62166	&	E	&	12 59 11.5	&	28 00 33	&	16.70	&	1.50	&	20.72	&	6895	&	3	&	48	&	3	&	20.3\\
4042	&	63244	&	SB0	&	12 58 48.1	&	28 01 07	&	17.03	&	1.64	&	21.91	&	7116	&	5	&	25	&	5	&	7.6\\
2879	&	68886	&	E	&	13 00 11.1	&	28 03 55	&	16.67	&	1.52	&	20.90	&	7350	&	4	&	46	&	3	&	20.4\\
3387	&	72587	&	E	&	12 59 31.6	&	28 06 02	&	17.29	&	1.42	&	21.92	&	7427	&	4	&	33	&	4	&	12.3\\
2976	&	79519	&	S0	&	13 00 04.2	&	28 09 18	&	16.63	&	1.49	&	20.81	&	6586	&	3	&	40	&	3	&	17.4\\
3699	&	81862	&	S0/Sp	&	12 59 09.9	&	28 09 52	&	17.36	&	1.35	&	21.57	&	8612	&	5	&	28	&	4	&	12.3\\
3204	&	82303	&	E	&	12 59 44.1	&	28 10 35	&	16.98	&	1.54	&	20.58	&	8316	&	3	&	42	&	3	&	18.1\\
4122	&	86791	&	dE	&	12 58 37.4	&	28 13 10	&	17.61	&	1.55	&	21.68	&	6784	&	5	&	31	&	5	&	9.2\\
3640	&	90411	&	E/S0	&	12 59 15.0	&	28 15 03	&	16.60	&	1.00	&	21.47	&	7428	&	4	&	23	&	3	&	21.2\\
3902	&	104960	&	dE	&	12 58 55.8	&	28 21 14	&	18.09	&	1.63	&	22.16	&	6355	&	6	&	15	&	5	&	6.7\\	

\end{longtable}
}
\bsp

\label{lastpage}
\end{document}